\documentclass[prl,twocolumn]{revtex4}
\usepackage{graphicx}
\usepackage{float}
\begin{document}

\title{Comment on ``Human Time-Frequency Acuity Beats the Fourier Uncertainty Principle"} 

\author{G.S. Thekkadath}
\author{Michael Spanner}
\affiliation{National Research Council of Canada, 100 Sussex Drive, Ottawa, ON, Canada K1A 0R6}

\begin{abstract}
In the initial article [Phys. Rev. Lett. 110, 044301 (2013), arXiv:1208.4611]
it was claimed that human hearing can beat the Fourier uncertainty principle.
In this Comment, we demonstrate that the experiment designed and implemented in
the original article was ill-chosen to test Fourier uncertainty in human
hearing.
\end{abstract}

\date{\today}

\maketitle

The Gabor limit \cite{gabor},
\begin{equation}\label{EqGabor}
	\Delta t \Delta f \geq \frac{1}{4\pi},
\end{equation}
refers to the lower bound on the product of the standard deviations (STD) in
time ($\Delta t$) and frequency ($\Delta f$) of an audio signal.  This limit is
a consequence of the Fourier uncertainty principle. In their Letter, Oppenheim
and Magnasco \cite{oppmag} claim that human hearing can surpass this limit.
They design an experiment which establishes psychological limens, $\delta t$
and $\delta f$, and show that their subjects can discriminate signals that beat
a limen-based uncertainty
\begin{equation} \label{EqLimen}
	\delta t\delta f \geq \frac{1}{4\pi}.
\end{equation}
The frequency and time limens used by the authors relate to the accuracy with
which human participants can distinguish small frequency and time shifts
present in a sequence of three test pulses.  The sequence in question is
referred to as "Task 5" in the paper.   It is our view that their experiment is
ill-chosen to test Fourier uncertainty.

\begin{figure}[t]
\includegraphics[width=0.45\textwidth]{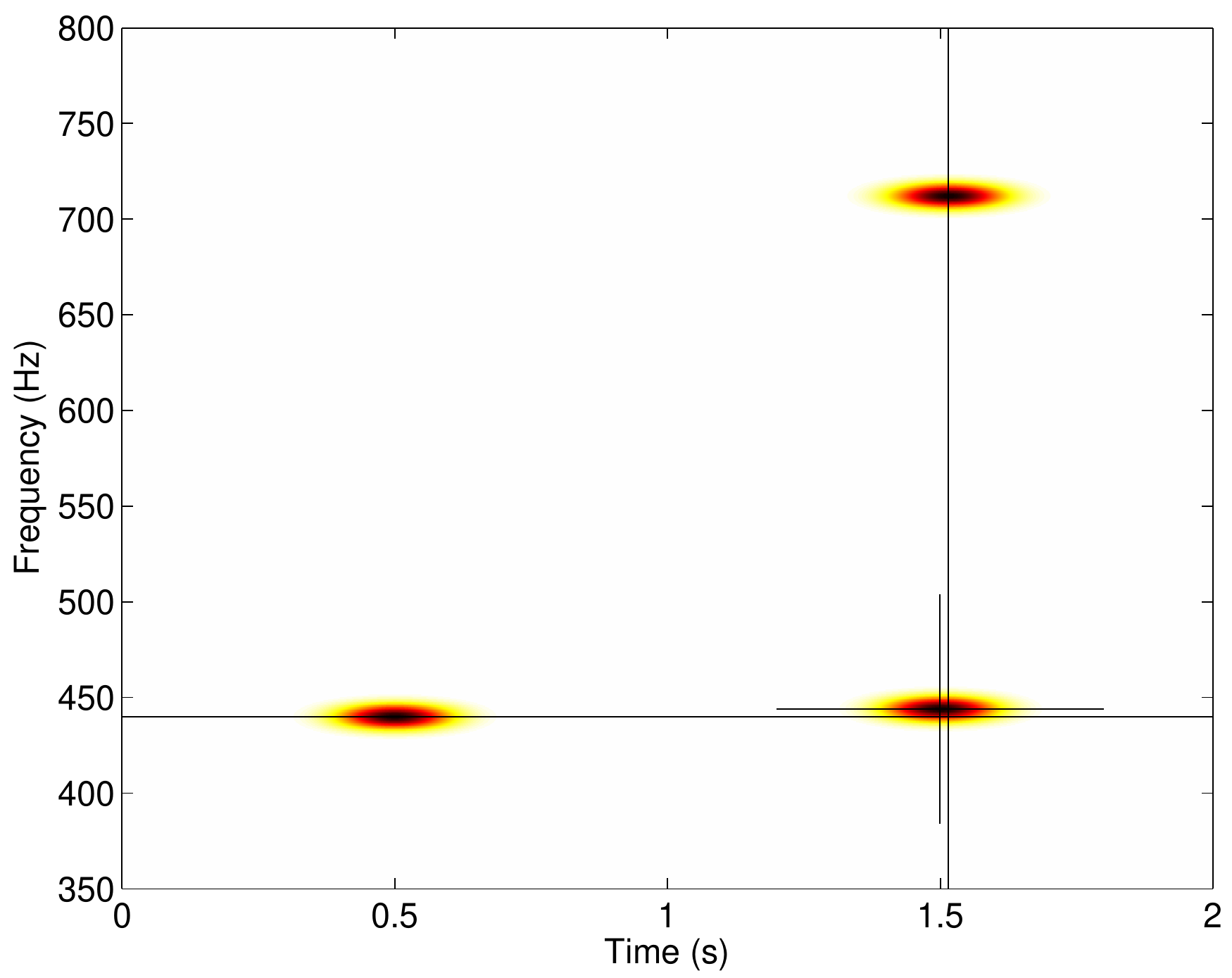}
	\caption[.]{$F(f,t_0)$ for the Gaussian
pulse train using the parameters $Dt$ = 0.015s, $Df$ = 4Hz, and $\sigma$ = 50 ms.} 
\end{figure}

Firstly, the $\Delta t$ and $\Delta f$ that appear in the Gabor limit must be
the STD of time and frequency {\it evaluated over the whole test signal}.  This
point is made clear in the derivation of the uncertainty principle that can be
found in the book of Cohen \cite{cohen} (Sections 3.2 and 3.3).  The limens
used by the authors, however, are simply {\it ad hoc} parameters that relate to
the STD of statistical errors made by the human participants when tasked with
estimating frequency and timing shifts in the test signal, and are unrelated to
the STD of time and frequency evaluated over the test signal.  Therefore, the
limen-based inequality Eq.(\ref{EqLimen}) is in no way related to the actual
Gabor limit Eq.(\ref{EqGabor}), and there is no expectation that the
limen-based inequality should be satisfied.

Secondly, one can straightforwardly use Fourier analysis itself to ``beat" Task
5, which again demonstrates that Task 5 does not test for violations of Fourier
uncertainty since any Fourier-based analysis would necessarily be limited by
the uncertainty principle.  One method is to use a window Fourier Transform
(WFT) to construct a spectrogram given by
\begin{equation} \label{EqWFT}
	F(f,t_0) = \int_{-\infty}^{\infty} \mathrm{e} ^ {i 2\pi f t} \mathrm{e} ^ {-\frac{(t-t_0)^2}{2\gamma^2}} X(t) \mathrm{d}t
\end{equation}
where $X(t)$ is the pulse sequence and $\gamma$ is the width of the window
function. The $\gamma$ is a free parameter that controls the aspect ratio of
the individual signals as they appear in $F(f,t_0)$, and must be chosen to
ensure that the signals do not overlap in $F(f,t_0)$.  This can be readily
achieved by setting $\gamma$ equal to, for example, the temporal variance of
the first pulse received in the pulse sequence.  In Fig.1, we show $F(f,t_0)$
for the sequence of Gaussian pulses used in Ref.\cite{oppmag}.  Fig.~1 clearly
demonstrates that the function $F(f,t_0)$ can be used to obtain all the
frequency and time shifts required to perform Task 5.  When evaluating the WFT
integral using a fast Fourier transform (FFT), the peak positions can be
resolved to (at least) within one grid point in both frequency ($df$) and time
($dt$).  For a sampling rate $R$ and time range $T$, the grid spacings in the
FFT are given by $dt = 1/R$ and $df=1/T$.  Therefore, the WFT would result in a
limen-based uncertainty of $\delta t\delta f = dtdf = 1/(RT)$.  Using $R$=96
kHz and $T$=2s, for example, we get $\delta t\delta f = dtdf =
5.2083\times10^{-6}$ which is orders of magnitude smaller than $1/(4\pi)
\approx 0.0796$, and also orders of magnitude smaller than what was achieved by
the human participants.  Furthermore, since the WFT is a linear transform that
can ``beat" Task 5, we also conclude that Task 5 does not
test for the necessity of nonlinear transforms in models of human hearing.\\

\end{document}